\documentclass[aps,prl,preprint,groupedaddress]{revtex4} 

\usepackage{graphicx}
\usepackage{bm}

\begin{document}


\title{Spectroscopic determination of the $s$-wave scattering lengths
of $^{86}$Sr and $^{88}$Sr}


\author{ P. G. Mickelson, Y. N. Martinez, A. D. Saenz, S. B. Nagel, Y. C. Chen,
and T. C. Killian}

\affiliation{Rice University, Department of Physics and Astronomy, Houston, Texas, 77251}

\author{P. Pellegrini, and R. C\^ot\'e}

\affiliation{Department of Physics, U-3046, University of Connecticut, Storrs, CT, 06269-3046}


\date{\today}

\begin{abstract}
We report the use of photoassociative spectroscopy to
determine the ground state
$s$-wave scattering lengths for the main bosonic isotopes of strontium,
$^{86}$Sr and $^{88}$Sr.
Photoassociative transitions are driven with a laser red-detuned
by up to 1400\,GHz from the $^1S_0$ - $^1P_1$ atomic resonance at
461\,nm. A minimum in the transition amplitude for $^{86}$Sr at
$-494 \pm 5$\,GHz allows us to determine the
scattering lengths $610$\,$a_0 < a_{86} < 2300$\,$a_0$ for $^{86}$Sr and
  a much  smaller value of
$-1$\,$a_0 < a_{88} < 13$\,$a_0$ for $^{88}$Sr.

\end{abstract}


\maketitle



    Photoassociative spectroscopy (PAS) of ultracold gases, in which a laser
    field resonantly excites colliding atoms to ro-vibrational states
    of excited molecular
    potentials, is a powerful probe of atomic cold
    collisions \cite{wbz99}. Transition frequencies have been used
    to obtain dispersion coefficients
    of molecular potentials, which yield the most accurate value
    of the atomic  excited-state lifetime\cite{mar95,jjl96,nms05}. Transition 
    amplitudes are related to the
    wave-function for colliding ground state atoms \cite{nww94,jul96},
    and
    can be used to determine the ground state $s$-wave
    scattering length \cite{gcm95,amg97,twj96,wtj99,dbw03,tkh04}.

    The $s$-wave scattering length is a crucial parameter
    for determining the efficiency of
    evaporative cooling and the stability of a
    Bose-Einstein condensate (BEC). It also
    sets the scale for collisional frequency shifts, which can limit the
    accuracy and stability of atomic frequency standards.

    The cold collision properties of
    alkaline-earth atoms like strontium, calcium, and magnesium,
    and atoms with similar electronic
    structure such as ytterbium, are currently
    the focus of intense study. These atoms
    posses narrow optical resonances that
    have great potential for optical frequency standards \cite{ktp03,tka03,obf99,sat05,hcn05}.
    Laser-cooling on narrow transitions is an efficient route to high
    phase-space density \cite{iik00,mki03}, and a BEC was
    recently produced with ytterbium \cite{tmk03}. Fundamental interest
    in alkaline-earth atoms is also high because their simple
    molecular potentials allow accurate tests of
    cold collision theory \cite{mjs01,ctk04,mna01}.

    PAS of calcium \cite{dbw03} and ytterbium \cite{tkh04}
    was recently used to
    determine $s$-wave scattering lengths of these atoms.
    This paper reports the use of PAS to determine the ground state
    $s$-wave scattering lengths for the main bosonic isotopes of strontium,
    $^{86}$Sr and $^{88}$Sr, which have relative abundances of 10\% and 83\%
    respectively. We find a huge scattering length for $^{86}$Sr
    of $610$\,$a_0 < a_{86} < 2300$\,$a_0$. Appreciable uncertainty comes from the value
of $C_6$ for the ground state potential.
In contrast, for $^{88}$Sr we find $-1$\,$a_0 < a_{88} < 13$\,$a_0$.
    From the data, we also make an improved measurement \cite{nms05} of the
$5s5p$\,${^1P_1}$ atomic lifetime ($\tau=5.25 \pm 0.01$\,ns).
PAS results for $^{88}$Sr, yielding a similar value of $\tau$ and a different value of $a_{88}$, were
    recently posted \cite{ykt05}.

      For PAS of strontium, atoms are initially trapped in a magneto-optical trap (MOT) operating on the
461\,nm  $^1S_0$ - $^1P_1$ transition, as described in
 \cite{nsl03,nms05}.
 We are able to produce pure samples of each isotope from the same atomic beam
 due to the intrinsic isotope selectivity of a MOT.
 For $^{86}$Sr and $^{88}$Sr respectively, about $7\times 10^7$ and $2.5\times 10^8$ atoms are
 trapped and cooled
 to $2$\,mK.

After this  stage,
the 461\,nm laser-cooling light is extinguished, the
field gradient is reduced to 
$0.1$ G/cm, and $689$ nm light for the $^1S_0$ - $^3P_1$
intercombination-line MOT \cite{kii99} is switched on.
This MOT consists of three retro-reflected beams, each with
a diameter of 2\,cm and intensity of $400-800$\,$\mu$W/cm$^2$.
Initially, the frequency of the laser-cooling light is  detuned
from
atomic resonance by about -1.3\,MHz, and spectrally broadened with a
$\pm$1.0\,MHz
sine-wave modulation.

During a 50\,ms transfer and equilibration period,
the field gradient and spectral modulation are linearly ramped to
$0.8$ G/cm and $\pm$0.7\,MHz.
The detuning is ramped to about -0.9\,MHz.
This yields about
$4\times 10^7$ $^{86}$Sr atoms at a temperature of $5\pm 1$\,$\mu$K
or  $1.5\times 10^8$ $^{88}$Sr
atoms at a temperature of about $8\pm 2$\,$\mu$K.
The peak density for both isotopes is about $2\times 10^{11}$\,cm$^{-3}$.
The intercombination-line MOT parameters are then held constant during
an adjustable hold time. 
In the absence of PAS, the lifetime of atoms
in the trap is about 500\,ms, limited
by background gas collisions.

The size of the atom cloud, the number of atoms, and thus the peak density
are primarily determined with absorption imaging using the
$^1S_0$ - $^1P_1$
transition.
We image along the direction of gravity, and transverse $1/\sqrt{e}$ density radii
are $\sigma=250$\,$\mu$m for $^{86}$Sr and $\sigma=400$\,$\mu$m for $^{88}$Sr.
To obtain information on the cloud dimension along gravity  \cite{lil04},
an additional camera monitors
fluorescence perpendicular to this direction. The cloud is
smaller by approximately a factor of two in this axis.


To excite photoassociative resonances,
a PAS laser, tuned to the red of the atomic $^1S_0$ - $^1P_1$ transition at
461 nm, is applied to the atoms during
hold times of $200-800$\,ms. When the PAS laser is tuned to a molecular
resonance, photoassociation provides a loss mechanism for the MOT,
decreasing the number of atoms.
The PAS laser has negligible effect on the atom cloud size
or temperature.
When the PAS light is tuned far from a molecular
resonance, it has no effect on the rate of loss from the trap.

PAS light is generated from an extended cavity diode laser at 922 nm
    using second-harmonic generation in a linear
   enhancement cavity \cite{bft97}.
    The laser linewidth is 80 MHz in the blue on a millisecond time scale,
    which   contributes significantly to the observed
    PAS resonance linewidths.
    The laser frequency is measured with a wavemeter that is regularly
    calibrated with the cooling laser whose frequency is locked to the
    atomic $^1S_0$ - $^1P_1$
    transition. The resulting 1-sigma statistical uncertainty for frequency measurements of the PAS
    laser is 200 MHz. There is also a comparable systematic uncertainty due to the sensitivity of the wavementer
    to the alignment of the laser into the device.


The available PAS laser power is as high as 20 mW. Several beam
geometries are used, but typically the PAS beam is retroreflected,
with a 1/e$^2$ intensity radius of about $w=1$\, mm, yielding a
maximum intensity of 3 W/cm$^2$ on the atoms.  Lower intensities
are obtained using an AOM. For some data, a quarter-wave plate is inserted in
the beam path after the first pass through the atoms to prevent
the formation of a standing wave. This variation has no noticeable affect on
the PAS intensity and frequency measurements.

 \begin{figure}
  \includegraphics[width=3.5in,clip=true, trim=00 155 0 0]{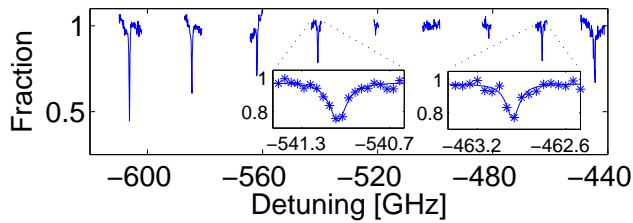}\\ 
  \caption{Selected region of the $^{86}$Sr PAS spectrum.
  The detuning is of the PAS laser with respect to the
  atomic $^1S_0$ - $^1P_1$ transition frequency. The vertical axis
  is the fraction of atoms remaining
  at the end of the hold time
  in comparison to the number in the absence of photoassociation.
  PAS laser intensity is 800 mW/cm$^2$, and
  hold times range from 350 to 450\,ms. The minimum of the transition
  amplitude occurs when the Condon radius corresponds to a zero of
  the ground state wave function. The insets show the quality of the data
  and the fits to Eqs.\ 1 and 2.
  }\label{86node}
\end{figure}

The observed PAS spectrum is relatively simple because the bosonic
isotopes of strontium lack hyperfine structure. In addition,
at the low temperatures of the intercombination-line
MOT, only $s$-wave collisions occur so only $J=1$ levels
are excited. Ground state $^1S_0$ atoms collide on a $^1\Sigma^+_g$ potential,
and,
of the four states converging to the $^1S_0$ + $^1P_1$ asymptote \cite{mjs01},
only the $^1\Sigma^+_u$ state 
is excited in this work.
Figure \ref{86node} shows  spectra recorded for
$^{86}$Sr in a region around 500 GHz to the red of the atomic transition.
The variation in transition amplitude will be discussed below, but
first we describe the method used to quantitatively
analyze the spectra.

The atom density varies as $\dot{n}=-\beta(I,f) n^2 -\Gamma n$, where
$f$ and $I$ are the laser frequency and intensity. This implies
the number of atoms in the trap as a function
of hold time, $t$,   follows
\begin{equation}\label{number}
   N(t)={N_0 \rm{e}^{-\Gamma t} \over 1+
   {n_0 \beta(I_{pk},f) \zeta \over 2\sqrt{2}\Gamma}(1-\rm{e}^{-\Gamma t})},
\end{equation}
where  $N_0$  and $n_0$ are the number and peak density,
respectively, at the beginning of the hold time. $I_{pk}$ is the
peak laser intensity, and $\zeta=w^2/(2\sigma^2+w^2)$ accounts for
the laser-atom overlap. We approximate the density distribution as
a Gaussian. The one-body loss rate ($\Gamma$) and $n_0$ are fixed
at values determined from independent measurements. Fit values of
$N_0$ agree well with independent measurements and are a
check of the method.

\begin{figure}
  \includegraphics[width=3.25in]{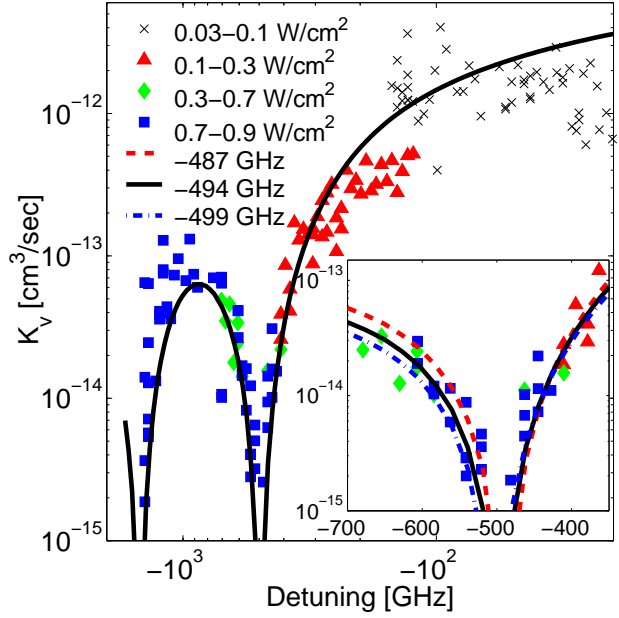}\\ 
  \caption{Experimental and theoretical values for $^{86}$Sr
photoassociative rate constants 
 for PAS laser
intensities of 1\,mW/cm$^2$ and a narrow laser linewidth.
  The experimental PAS laser intensity used
  for each measurement is indicated by the symbol. Theory is
  for 5\,$\mu$K atoms with $C_6=3170$\,a.u.
  (Inset) Experimental amplitudes are scaled by 1/1.4, which is reasonable
  given systematic uncertainties.
  The detuning of the minimum is used to determine the ground
  state potential as described in the text.
  The solid line is the best fit, corresponding to
  a minimum at -494\,GHz.
 The dashed and dash-dot lines correspond to
   -487\,GHz
   and -499\,GHz  respectively.
  }\label{86beta}
\end{figure}

\begin{figure}
  \includegraphics[width=3.5in, clip=true,trim=00 160 0 0]{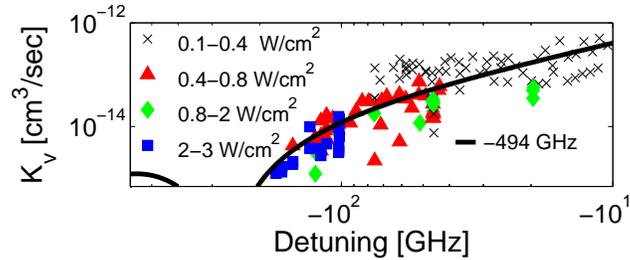}\\ 
  \caption{Same as Fig.\ \ref{86beta}, but for $^{88}$Sr. The theoretical curve, calculated for
  10\,$\mu$K atoms, is
  found using the potential determined from $^{86}$Sr with $C_6=3170$\,a.u.
  It predicts a scattering length of $a_{88}=6$\,$a_0$.
}\label{88beta}
\end{figure}

The photoassociative two-body loss rate,
$\beta$, near resonance $v$, is
approximated as
\begin{equation}\label{beta}
  \beta(I_{pk},f)={2 K_v I_{pk} \gamma_v\over \gamma}{1 \over 1 + 4(f-f_v)^2/\gamma^2},
\end{equation}
where $f_v$ is the center frequency for the transition.
The
experimental linewidth, $\gamma \approx 150$\,MHz,
approximately equals the sum of the natural radiative linewidth for the
transition, $\gamma_v=61$\,MHz \cite{nms05}, and the measured laser linewidth.
Including the factor $\gamma_v / \gamma$ in Eq.\ \ref{beta} 
accounts for broadening beyond $\gamma_v$\cite{mdh02,ssd02}. 
$I_{pk}$ is in units of mW/cm$^2$, and $K_v$
is the collision rate constant,
 on resonance,
for 1\,mW/cm$^2$ intensity from an ideal laser with negligible
linewidth. 
Difficulty in accurately determining the spatial dimensions of the
atom cloud and in aligning the PAS laser beam on the atoms leads
to systematic uncertainties of about a factor of two in
measurements of $K_v$ (Figs.\ \ref{86beta} and \ref{88beta}).
Except for the most intense PAS transitions close to atomic
resonance for $^{86}$Sr,
the linear variation with intensity used
in Eq.\ \ref{beta} is a good approximation and
$K_v$ is a constant for each transition.

From the observed transition frequencies we can obtain an accurate value of
the atomic $5s5p$\,$^1P_1$ lifetime, $\tau$. For the region of the
 molecular potentials probed
by PAS, the transition energies can be described with the semiclassical approximation \cite{lbe70}
\begin{eqnarray}\label{potential} \label{eq:V(R)}
   E(v)&=& D-X_0(v_D-v)^6, \nonumber\\
  X_0 &=&\left[{\Gamma(4/3) \over 2\sqrt{2\pi}\Gamma(5/6)}\right]^6{h^6 \over \mu^3C_3^2},\,\,\,
\end{eqnarray}
where $D$ is the  dissociation energy, $\mu$ the reduced mass, 
$\lambda = 461$ nm, $C_3 = 3 \hbar \lambda^3/ 16 \pi^3 \tau$, and
$v_D$ is a non-integer quantum number corresponding to a hypothetical level at the dissociation limit.
Relativistic retardation, rotational energy, higher order dispersion terms, ground state potential curvature,
thermal shifts, and all light-induced shifts are negligible at our level of accuracy.
A fit to the $^{86}$Sr data yields $\tau=5.25\pm0.01$\,ns.
The $^{88}$Sr data is not extensive enough to contribute
to the determination.

While all the data in
Figs.\ \ref{86beta} and \ref{88beta} are valuable, the
 detunings of the laser corresponding to minima of the PAS rate are particularly important.
At these
detunings,
the Condon radius for the excitation matches a node of the
ground-state wave function, which is equivalent to saying the overlap integral between the ground-
and excited-state wavefunctions vanishes.
Minima positions provide precise enough information
about ground-state potentials and wavefunctions to determine
$s$-wave scattering
lengths to high accuracy.
The  minima positions
are also independent of atom density and laser intensity calibrations, and
 for the ultracold gases used here, they are independent of
$T$, the temperature of the atoms.

For excitation  by a laser with negligible linewidth, the
collision rate constant
at ultra-low temperatures and on resonance is \cite{nww94}
\begin{equation}
    K_{v}(T,I)=\frac{1}{hQ_T}\int_0^{\infty}
    d\epsilon e^{-\epsilon/k_BT}\frac{\gamma_v \gamma_s(\epsilon,J=0)}{(\epsilon^2 + (\gamma/2)^2)},
\end{equation}
where $Q_T=(2\pi\mu k_BT/h^2)^{3/2}$, $k_B$ is the Boltzmann constant,
  $\gamma=\gamma_v+\gamma_s(\epsilon,J=0)$ with
  $\gamma_s(\epsilon,J=0)$ equal to the laser-stimulated width.
   At low laser intensities,  $\gamma_s(\epsilon,J=0)$ can be expressed using  Fermi's golden rule
 as $\gamma_s(\epsilon,J=0)=\pi I d^2/\epsilon_0 c $. Here, $\epsilon_0$ is
  the vacuum permittivity, $c$ the
  vacuum speed of light and $d^2=|\langle v|{D({R})}|\epsilon \rangle |^2$ where ${D({R})}$ is the molecular
  dipole transition moment connecting $|v \rangle$ and $|\epsilon \rangle$, the excited vibrational wave function and the
  energy normalized ground continuum wave
  function respectively. Because
 only bound levels close to the
  potential dissociation limit are excited, $\gamma_v$ is
  independent of  $v$ and equal to twice the atomic linewidth $\gamma_{at}$.
  By the same argument, ${D({R})}$ can be approximated as independent of $R$ \cite{cks03}
  and is connected to the $^1S_0$ - $^1P_1$
  atomic dipole moment via ${D({R})}=\sqrt{2}\alpha d_{at}$. Using
  $d_{at}=\sqrt{3\pi \epsilon_0 \hbar \gamma_{at}\lambdabar^3}$ and the line strength factor
  $\alpha=\sqrt{2/3}$ \cite{mjs01},
  we find $D=3.6$ a.u.


    For our analysis, the inner part ($R<19$\,$a_0$) of ground- and excited-state potentials are formed
by  an experimental RKR potential
\cite{gms84} and  ab initio potential \cite{baa96} respectively. These are
smoothly connected
 to the multipolar van der Waals expansion in $C_n/R^n$ at large
internuclear separation. For the excited-state potential, only the $C_3$
term contributes, and we use the value determined above. For the
ground-state potential, $C_6$, $C_8$ and $C_{10}$ terms are
included \cite{pde02,mbr03,c6note}.
Relativistic retardation effects  in the asymptotic part of the excited-state
 potential are treated as described in \cite{nms05}. Wave functions are calculated using a full quantum calculation.
 For the bound vibrational
 levels, the Mapped Fourier Grid Method \cite{kdk99} has been used whereas the ground-state continuum wave function was calculated
 using a Numerov algorithm.

Large transition moments for $^{86}$Sr allow us to experimentally
characterize about 80 transitions for this isotope, extending to
detunings as large as -1400\,GHz (Fig.\ \ref{86beta}). This allows
us to clearly identify minima in the transition moment at
$-494\pm 5$\,GHz and $-1500$ GHz,  corresponding to Condon radii 
of $62.6\pm 0.2 $\,$a_0$ and $43 $\,$a_0$  respectively. 
To precisely determine the position of the  minimum at
$-494$\,GHz, we calculated a set of potential curves with
different node positions by varying the position of the
 repulsive inner-wall. Using these curves we performed a least-squares fit to the data shown in the inset of Fig.\
 \ref{86beta}. Quoted uncertainties are one standard deviation.
 An overall amplitude factor was also varied but the
 best values were always well within the experimental uncertainties discussed
 above.
 The determined position of the node was independent of
 the value of the ground state $C_6$ coefficient used. It also did
 not change significantly if only data from -600 to -400 GHz was
 fit.


 The Sr ground state potential determined from the node position and the ground state
 $C_6$ coefficient accurately describe the
 collisional properties of the system. This allows us to
 determine the $s$-wave scattering length $a_{86}$ from the
 calculated zero-energy continuum wave function. However, we must carefully address
 the uncertainty in $a_{86}$ arising from uncertainty in $C_6$. The best values in the
 literature are a semiempirical method yielding
 $C_6=3250$\, a.u. \cite{mbr03}, and a
 relativistic many-body calculation 
 of $C_6=3170$\, a.u. \cite{pde02}, although a reevaluation of \cite{pde02} inputting
 a more recent value of the $5s5p$\,$^1P_1$ lifetime \cite{ykt05}
 predicts $C_6=3103(7)$\, a.u. \cite{c6note}.
If we take the latter value, and a minimum at $-494 \pm 5$\, GHz, we find
 $a_{86}=1450^{+850}_{-370}$\,$a_0$.
 ($a_{86}$
would diverge for
a minimum position of $-482$\,GHz.) This large scattering length indicates a bound level in
 the ground state potential very near
 threshold, which implies that $a_{86}$ is also very sensitive to $C_6$.
 For $C_6=3170$\, a.u.  and $C_6=3240$\, a.u.
we calculate $a_{86}=940^{+279}_{-170}$\,$a_0$ and
$a_{86}=700^{+140}_{-90}$\,$a_0$ respectively. The spread in $C_6$
values makes it difficult to quote a rigorous statistical best
value and uncertainty  for $a_{86}$. Taking the extremes of the
values found above, including spread in $C_6$ and one-standard-deviation variation in node position,
we find $610$\,$a_0 < a_{86} < 2300$\,$a_0$.



   PAS transitions in $^{88}$Sr are significantly weaker than in $^{86}$Sr,
   and only 50 transitions were characterized, extending to binding energies of
   -174\,GHz (Fig.\ \ref{88beta}). We lack the sensitivity required to observe
transitions to the red of
   the first minimum in the transition amplitude and thus cannot independently
   determine its position
   with high accuracy. However, the potential  found using $^{86}$Sr should
   also determine the wavefunctions for $^{88}$Sr.
   Because of the heavy mass of strontium, effects due to the breakdown of the
    Born-Oppenheimer approximation
    should be
    negligible as is the case for rubidium {\cite{kkh02}.
Predicted photoassociation rates agree with measurements within
    experimental uncertainties (Fig.\ \ref{88beta}), and
    we calculate $a_{88}=10 \pm 3$\,a$_0$ and a wave-function node at
    $76 \pm 0.4$\,a$_0$ for $C_6=3103$\, a.u.
    For $C_6=3240$\, a.u., $a_{88}=2 \pm 3$\,a$_0$. This yields
    $-1$\,a$_0 < a_{88} <13$\,a$_0$.

In conclusion,  we determined the scattering
lengths of $^{86}$Sr and $^{88}$Sr from PAS spectra. Our
  analysis uses full quantum calculations
 and is not based on semi-classical
 arguments \cite{bav00,ykt05} that can be invalid if the scattering length becomes
 divergent (as in $^{86}$Sr), small (as in $^{88}$Sr), or negative.
The large positive value for $^{86}$Sr is very promising for the realization
of a BEC through evaporative cooling in an optical trap.
Such a condensate would have the advantage that losses due
to inelastic processes would be minimal since there is no hyperfine interaction
and spin-exhange collisions would not take place. The lifetime of a
$^{86}$Sr condensate would be influenced mostly by three-body recombination.
BEC of $^{88}$Sr will be a significant challenge \cite{iik00}
because of the small
scattering length, but it may
allow studies of a very weakly interacting quantum gas with behavior similar to a
hydrogen condensate \cite{fkw98}.

This research was supported by the Welch Foundation (Grant \#
C-1579), Office for Naval Research, and David and
Lucille Packard Foundation. The authors are grateful to A. R.
Allouche for providing them the Sr$_2$ ab initio potentials,and to
E. Tiesinga and P. Julienne for valuable discussions.


\begin{thebibliography}{38}
\expandafter\ifx\csname natexlab\endcsname\relax\def\natexlab#1{#1}\fi
\expandafter\ifx\csname bibnamefont\endcsname\relax
  \def\bibnamefont#1{#1}\fi
\expandafter\ifx\csname bibfnamefont\endcsname\relax
  \def\bibfnamefont#1{#1}\fi
\expandafter\ifx\csname citenamefont\endcsname\relax
  \def\citenamefont#1{#1}\fi
\expandafter\ifx\csname url\endcsname\relax
  \def\url#1{\texttt{#1}}\fi
\expandafter\ifx\csname urlprefix\endcsname\relax\def\urlprefix{URL }\fi
\providecommand{\bibinfo}[2]{#2}
\providecommand{\eprint}[2][]{\url{#2}}

\bibitem[{\citenamefont{Weiner et~al.}(1999)\citenamefont{Weiner, Bagnato,
  Zilio, and Julienne}}]{wbz99}
\bibinfo{author}{\bibfnamefont{J.}~\bibnamefont{Weiner} \bibnamefont{{\it et al.}}},
  \bibinfo{journal}{{ Rev. Mod. Phys.}} \textbf{\bibinfo{volume}{71}},
  \bibinfo{pages}{1} (\bibinfo{year}{1999}).

\bibitem[{\citenamefont{McAlexander et~al.}(1995)\citenamefont{McAlexander,
  Abraham, Ritchie, Williams, Stoof, and Hulet}}]{mar95}
\bibinfo{author}{\bibfnamefont{W.~I.} \bibnamefont{McAlexander} \bibnamefont{{\it et al.}}},
  \bibinfo{journal}{{ Phys. Rev. A}} \textbf{\bibinfo{volume}{51}},
  \bibinfo{pages}{R871} (\bibinfo{year}{1995}).

\bibitem[{\citenamefont{Jones et~al.}(1996)\citenamefont{Jones, Julienne, Lett,
  Phillips, Tiesinga, and Williams}}]{jjl96}
\bibinfo{author}{\bibfnamefont{K.~M.} \bibnamefont{Jones} \bibnamefont{{\it et al.}}},
  \bibinfo{journal}{{ Europhys. Lett.}} \textbf{\bibinfo{volume}{35}},
  \bibinfo{pages}{85} (\bibinfo{year}{1996}).

\bibitem[{\citenamefont{Nagel et~al.}(2005)\citenamefont{Nagel, Mickelson,
  Saenz, Martinez, Chen, Killian, Pellegrini, and C\^ot\'e}}]{nms05}
\bibinfo{author}{\bibfnamefont{S.~B.} \bibnamefont{Nagel} \bibnamefont{{\it et al.}}},
  \bibinfo{journal}{{ Phys. Rev. Lett.}} \textbf{\bibinfo{volume}{94}},
  \bibinfo{pages}{083004} (\bibinfo{year}{2005}).

\bibitem[{\citenamefont{Napolitano et~al.}(1994)\citenamefont{Napolitano,
  Weiner, Williams, and Julienne}}]{nww94}
\bibinfo{author}{\bibfnamefont{R.}~\bibnamefont{Napolitano} \bibnamefont{{\it et al.}}},
  \bibinfo{journal}{{ Phys. Rev. Lett}}
  \textbf{\bibinfo{volume}{73}}, \bibinfo{pages}{1352} (\bibinfo{year}{1994}).

\bibitem[{\citenamefont{Julienne}(1996)}]{jul96}
\bibinfo{author}{\bibfnamefont{P.}~\bibnamefont{Julienne}}, \bibinfo{journal}{{
  J. Res. Natl. Inst. Stand. Technol.}} \textbf{\bibinfo{volume}{101}},
  \bibinfo{pages}{487} (\bibinfo{year}{1996}).

\bibitem[{\citenamefont{Gardner et~al.}(1995)\citenamefont{Gardner, Cline,
  Miller, Heinzen, Boesten, and Verhaar}}]{gcm95}
\bibinfo{author}{\bibfnamefont{J.~R.} \bibnamefont{Gardner} \bibnamefont{{\it et al.}}},
  \bibinfo{journal}{{ Phys. Rev. Lett.}}
  \textbf{\bibinfo{volume}{74}}, \bibinfo{pages}{3764} (\bibinfo{year}{1995}).

\bibitem[{\citenamefont{Abraham et~al.}(1997)\citenamefont{Abraham,
  McAlexander, Gerton, Hulet, C\^{o}t\'{e}, and Dalgarno}}]{amg97}
\bibinfo{author}{\bibfnamefont{E.~R.~I.} \bibnamefont{Abraham} \bibnamefont{{\it et al.}}},
  \bibinfo{journal}{{Phys. Rev. A}} \textbf{\bibinfo{volume}{55}},
  \bibinfo{pages}{R3299} (\bibinfo{year}{1997}).

\bibitem[{\citenamefont{Tiesinga et~al.}(1996)\citenamefont{Tiesinga, Williams,
  Julienne, Jones, Lett, and Phillips}}]{twj96}
\bibinfo{author}{\bibfnamefont{E.}~\bibnamefont{Tiesinga} \bibnamefont{{\it et al.}}},
  \bibinfo{journal}{{J. Res. Natl. Inst. Stand. Technol.}}
  \textbf{\bibinfo{volume}{101}}, \bibinfo{pages}{505} (\bibinfo{year}{1996}).

\bibitem[{\citenamefont{Williams et~al.}(1999)\citenamefont{Williams, Tiesinga,
  Julienne, Wang, Stwalley, and Gould}}]{wtj99}
\bibinfo{author}{\bibfnamefont{C.~J.} \bibnamefont{Williams} \bibnamefont{{\it et al.}}},
  \bibinfo{journal}{{Phys. Rev. A}} \textbf{\bibinfo{volume}{60}},
  \bibinfo{pages}{4427} (\bibinfo{year}{1999}).

\bibitem[{\citenamefont{Degenhardt et~al.}(2003)\citenamefont{Degenhardt,
  Binnewies, Wilpers, Sterr, Riehle, Lisdat, and Tiemann}}]{dbw03}
\bibinfo{author}{\bibfnamefont{C.}~\bibnamefont{Degenhardt} \bibnamefont{{\it et al.}}},
   \bibinfo{journal}{{ Phys. Rev. A}} \textbf{\bibinfo{volume}{67}},
  \bibinfo{pages}{043408} (\bibinfo{year}{2003}).

\bibitem[{\citenamefont{Takasu et~al.}(2004)\citenamefont{Takasu, Komori,
  Honda, Kumakura, Yabuzaki, and Takahashi}}]{tkh04}
\bibinfo{author}{\bibfnamefont{Y.}~\bibnamefont{Takasu} \bibnamefont{{\it et al.}}},
  \bibinfo{journal}{{ Phys. Rev. Lett}} \textbf{\bibinfo{volume}{93}},
  \bibinfo{pages}{123202} (\bibinfo{year}{2004}).

\bibitem[{\citenamefont{Katori et~al.}(2003)\citenamefont{Katori, Takamoto,
  Pal'chikov, and Ovsiannikov}}]{ktp03}
\bibinfo{author}{\bibfnamefont{H.}~\bibnamefont{Katori} \bibnamefont{{\it et al.}}},
  \bibinfo{journal}{{ Phys. Rev. Lett.}}
  \textbf{\bibinfo{volume}{91}}, \bibinfo{pages}{173005}
  (\bibinfo{year}{2003}).

\bibitem[{\citenamefont{Takamoto and Katori}(2003)}]{tka03}
\bibinfo{author}{\bibfnamefont{M.}~\bibnamefont{Takamoto}} \bibnamefont{and}
  \bibinfo{author}{\bibfnamefont{H.}~\bibnamefont{Katori}}, \bibinfo{journal}{{
  Phys. Rev. Lett.}} \textbf{\bibinfo{volume}{91}}, \bibinfo{pages}{223001}
  (\bibinfo{year}{2003}).

\bibitem[{\citenamefont{Oates et~al.}(1999)\citenamefont{Oates, Bondu, Fox, and
  Hollberg}}]{obf99}
\bibinfo{author}{\bibfnamefont{C.~W.} \bibnamefont{Oates} \bibnamefont{{\it et al.}}},
  \bibinfo{journal}{{ Eur. Phys. J. D}} \textbf{\bibinfo{volume}{7}},
  \bibinfo{pages}{449} (\bibinfo{year}{1999}).

\bibitem[{\citenamefont{Santra et~al.}(2004)\citenamefont{Santra, Arimondo,
  Ido, Greene, and Ye}}]{sat05}
\bibinfo{author}{\bibfnamefont{R.}~\bibnamefont{Santra} \bibnamefont{{\it et al.}}},
  \bibinfo{journal}{{physics/0411197}}  (\bibinfo{year}{2004}).

\bibitem[{\citenamefont{Hong et~al.}(2004)\citenamefont{Hong, Cramer,
  Nagourney, and Fortson}}]{hcn05}
\bibinfo{author}{\bibfnamefont{T.}~\bibnamefont{Hong} \bibnamefont{{\it et al.}}},
 \bibinfo{journal}{{
  Phys. Rev. Lett.}} \textbf{\bibinfo{volume}{94}}, \bibinfo{pages}{50801}
  (\bibinfo{year}{2005}).

\bibitem[{\citenamefont{Ido et~al.}(2000)\citenamefont{Ido, Isoya, and
  Katori}}]{iik00}
\bibinfo{author}{\bibfnamefont{T.}~\bibnamefont{Ido} \bibnamefont{{\it et al.}}},
  \bibinfo{journal}{{
  Phys. Rev. A}} \textbf{\bibinfo{volume}{61}}, \bibinfo{pages}{061403(R)}
  (\bibinfo{year}{2000}).

\bibitem[{\citenamefont{Mukaiyama et~al.}(2003)\citenamefont{Mukaiyama, Katori,
  Ido, Li, and Kuwata-Gonokami}}]{mki03}
\bibinfo{author}{\bibfnamefont{T.}~\bibnamefont{Mukaiyama} \bibnamefont{{\it et al.}}},
    \bibinfo{journal}{{ Phys. Rev. Lett.}} \textbf{\bibinfo{volume}{90}},
  \bibinfo{pages}{113002} (\bibinfo{year}{2003}).

\bibitem[{\citenamefont{Takasu et~al.}(2003)\citenamefont{Takasu, Maki, Komori,
  Takano, Honda, Kumakura, Yabuzaki, and Takahashi}}]{tmk03}
\bibinfo{author}{\bibfnamefont{Y.}~\bibnamefont{Takasu} \bibnamefont{{\it et al.}}},
   \bibinfo{journal}{{ Phys. Rev. Lett.}} \textbf{\bibinfo{volume}{91}},
  \bibinfo{pages}{040404} (\bibinfo{year}{2003}).

\bibitem[{\citenamefont{Machholm et~al.}(2001)\citenamefont{Machholm, Julienne,
  and Suominen}}]{mjs01}
\bibinfo{author}{\bibfnamefont{M.}~\bibnamefont{Machholm}},
  \bibinfo{author}{\bibfnamefont{P.~S.} \bibnamefont{Julienne}},
  \bibnamefont{and} \bibinfo{author}{\bibfnamefont{K.-A.}
  \bibnamefont{Suominen}}, \bibinfo{journal}{{ Phys. Rev. A}}
  \textbf{\bibinfo{volume}{64}}, \bibinfo{pages}{033425}
  (\bibinfo{year}{2001}).

\bibitem[{\citenamefont{Ciurylo et~al.}(2004)\citenamefont{Ciurylo, Tiesinga,
  Kotochigova, and Julienne}}]{ctk04}
\bibinfo{author}{\bibfnamefont{R.}~\bibnamefont{Ciurylo} \bibnamefont{{\it et al.}}},
 \bibinfo{journal}{{ Phys. Rev. A}}
  \textbf{\bibinfo{volume}{70}}, \bibinfo{pages}{062710}
  (\bibinfo{year}{2004}).

\bibitem[{\citenamefont{Montalv{\~a}o and de~Jesus~Napolitano}(2001)}]{mna01}
\bibinfo{author}{\bibfnamefont{R.~W.} \bibnamefont{Montalv{\~a}o}}
  \bibnamefont{and}
  \bibinfo{author}{\bibfnamefont{R.J.}~\bibnamefont{Napolitano}},
  \bibinfo{journal}{{ Phys. Rev. A}} \textbf{\bibinfo{volume}{64}},
  \bibinfo{pages}{011403(R)} (\bibinfo{year}{2001}).

 \bibitem[{\citenamefont{Yasuda et~al.}(2005)\citenamefont{Yasuda, Kishimoto,
  Takamoto, and Katori}}]{ykt05}
\bibinfo{author}{\bibfnamefont{M.}~\bibnamefont{Yasuda}} \bibnamefont{{\it et al.}},
  \bibinfo{journal}{physics/0501053}  (\bibinfo{year}{2005}).


\bibitem[{\citenamefont{Nagel et~al.}(2003)\citenamefont{Nagel, Simien, Laha,
  Gupta, Ashoka, and Killian}}]{nsl03}
\bibinfo{author}{\bibfnamefont{S.~B.} \bibnamefont{Nagel} \bibnamefont{{\it et al.}}},
 \bibinfo{journal}{{ Phys. Rev. A}}
  \textbf{\bibinfo{volume}{67}}, \bibinfo{pages}{011401(R)}
  (\bibinfo{year}{2003}).

\bibitem[{\citenamefont{Katori et~al.}(1999)\citenamefont{Katori, Ido, Isoya,
  and Kuwata-Gonokami}}]{kii99}
\bibinfo{author}{\bibfnamefont{H.}~\bibnamefont{Katori} \bibnamefont{{\it et al.}}},
  \bibinfo{journal}{{ Phys. Rev. Lett.}} \textbf{\bibinfo{volume}{82}},
  \bibinfo{pages}{1116} (\bibinfo{year}{1999}).

\bibitem[{\citenamefont{Loftus et~al.}(2004)\citenamefont{Loftus, Ido, Ludlow,
  Boyd, and Ye}}]{lil04}
\bibinfo{author}{\bibfnamefont{T.}~\bibnamefont{Loftus} \bibnamefont{{\it et al.}}},
  \bibinfo{journal}{{
  Phys. Rev. Lett.}} \textbf{\bibinfo{volume}{93}}, \bibinfo{pages}{073003}
  (\bibinfo{year}{2004}).

\bibitem[{\citenamefont{Bode et~al.}(1997)\citenamefont{Bode, Freitag,
  Tunnermann, and Welling}}]{bft97}
\bibinfo{author}{\bibfnamefont{M.}~\bibnamefont{Bode} \bibnamefont{{\it et al.}}},
   \bibinfo{journal}{{ Opt. Lett.}} \textbf{\bibinfo{volume}{22}},
  \bibinfo{pages}{1220} (\bibinfo{year}{1997}).

\bibitem[{\citenamefont{McKenzie et~al.}(2002)\citenamefont{McKenzie,
  Denschlag, Häffner, Browaeys, de~Araujo, Fatemi, Jones, Simsarian, Cho,
  Simoni et~al.}}]{mdh02}
\bibinfo{author}{\bibfnamefont{C.}~\bibnamefont{McKenzie} \bibnamefont{{\it et al.}}},
 \bibinfo{journal}{{ Phys. Rev. Lett.}}
  \textbf{\bibinfo{volume}{88}}, \bibinfo{pages}{120403}
  (\bibinfo{year}{2002}).

\bibitem[{\citenamefont{Schloder et~al.}(2002)\citenamefont{Schloder, Silber,
  Deuschle, and Zimmermann}}]{ssd02}
\bibinfo{author}{\bibfnamefont{U.}~\bibnamefont{Schloder} \bibnamefont{{\it et al.}}},
  \bibinfo{journal}{{ Phys. Rev. A}} \textbf{\bibinfo{volume}{66}},
  \bibinfo{pages}{061403(R)} (\bibinfo{year}{2002}).

\bibitem[{\citenamefont{Leroy and B.Bernstein}(1970)}]{lbe70}
\bibinfo{author}{\bibfnamefont{R.~J.} \bibnamefont{Leroy}} \bibnamefont{and}
  \bibinfo{author}{\bibfnamefont{R.}~\bibnamefont{B.Bernstein}},
  \bibinfo{journal}{{J. Chem. Phys.}} \textbf{\bibinfo{volume}{52}},
  \bibinfo{pages}{3869} (\bibinfo{year}{1970}).

\bibitem[{\citenamefont{Czuchaj et~al.}(2003)\citenamefont{Czuchaj,
  Kro\'{s}nicki, and Stoll}}]{cks03}
\bibinfo{author}{\bibfnamefont{E.}~\bibnamefont{Czuchaj}},
  \bibinfo{author}{\bibfnamefont{M.}~\bibnamefont{Kro\'{s}nicki}},
  \bibnamefont{and} \bibinfo{author}{\bibfnamefont{H.}~\bibnamefont{Stoll}},
  \bibinfo{journal}{{Chem. Phys. Lett.}} \textbf{\bibinfo{volume}{371}},
  \bibinfo{pages}{401} (\bibinfo{year}{2003}).

\bibitem[{\citenamefont{Gerber et~al.}(1984)\citenamefont{Gerber, M{\"{o}}ller,
  and Schneider}}]{gms84}
\bibinfo{author}{\bibfnamefont{G.}~\bibnamefont{Gerber}},
  \bibinfo{author}{\bibfnamefont{R.}~\bibnamefont{M{\"{o}}ller}},
  \bibnamefont{and}
  \bibinfo{author}{\bibfnamefont{H.}~\bibnamefont{Schneider}},
  \bibinfo{journal}{{J. Chem. Phys.}} \textbf{\bibinfo{volume}{81}},
  \bibinfo{pages}{1538} (\bibinfo{year}{1984}).


\bibitem[{\citenamefont{Boutassetta et~al.}(1996)\citenamefont{Boutassetta,
  Allouche, and Aubert-Fr\'{e}con}}]{baa96}
\bibinfo{author}{\bibfnamefont{N.}~\bibnamefont{Boutassetta}},
  \bibinfo{author}{\bibfnamefont{A.~R.} \bibnamefont{Allouche}},
  \bibnamefont{and}
  \bibinfo{author}{\bibfnamefont{M.}~\bibnamefont{Aubert-Fr\'{e}con}},
  \bibinfo{journal}{{ Phys. Rev. A}} \textbf{\bibinfo{volume}{53}},
  \bibinfo{pages}{3845} (\bibinfo{year}{1996}).


\bibitem[{\citenamefont{Porsev and Derevianko}(2002)}]{pde02}
\bibinfo{author}{\bibfnamefont{S.~G.} \bibnamefont{Porsev}} \bibnamefont{and}
  \bibinfo{author}{\bibfnamefont{A.}~\bibnamefont{Derevianko}},
  \bibinfo{journal}{{ Phys. Rev. A}} \textbf{\bibinfo{volume}{65}},
  \bibinfo{pages}{020701(R)} (\bibinfo{year}{2002}).

\bibitem[{\citenamefont{Mitroy and Bromley}(2003)}]{mbr03}
\bibinfo{author}{\bibfnamefont{J.}~\bibnamefont{Mitroy}} \bibnamefont{and}
  \bibinfo{author}{\bibfnamefont{M.~W.~J.} \bibnamefont{Bromley}},
  \bibinfo{journal}{{ Phys. Rev. A}} \textbf{\bibinfo{volume}{68}},
  \bibinfo{pages}{052714} (\bibinfo{year}{2003}).

\bibitem[{c6n()}]{c6note}
\bibinfo{note}{
Private communication, A. Derevianko (2005).}

\bibitem[{\citenamefont{Kokoouline et~al.}(1999)\citenamefont{Kokoouline,
  Dulieu, Kosloff, and Masnou-Seeuws}}]{kdk99}
\bibinfo{author}{\bibfnamefont{V.}~\bibnamefont{Kokoouline} \bibnamefont{{\it et al.}}},
  \bibinfo{journal}{{J. Chem. Phys.}} \textbf{\bibinfo{volume}{110}},
  \bibinfo{pages}{9865} (\bibinfo{year}{1999}).

\bibitem[{\citenamefont{van Kempen et~al.}(2002)\citenamefont{van Kempen,
  Kokkelmans, Heinzen, and Verhaar}}]{kkh02}
\bibinfo{author}{\bibfnamefont{E.~G.~M.} \bibnamefont{van Kempen} \bibnamefont{{\it et al.}}},
  \bibinfo{journal}{{ Phys. Rev. Lett.}}
  \textbf{\bibinfo{volume}{88}}, \bibinfo{pages}{093201}
  (\bibinfo{year}{2002}).

\bibitem[{\citenamefont{Boisseau et~al.}(2000)\citenamefont{Boisseau, Audouard,
  Vigu\'{e}, and Julienne}}]{bav00}
\bibinfo{author}{\bibfnamefont{C.}~\bibnamefont{Boisseau} \bibnamefont{{\it et al.}}},
  \bibinfo{journal}{{Phys. Rev. A}}
  \textbf{\bibinfo{volume}{62}}, \bibinfo{pages}{052705}
  (\bibinfo{year}{2000}).


\bibitem[{\citenamefont{Fried et~al.}(1998)\citenamefont{Fried, Killian,
  Willmann, Landhuis, Moss, Kleppner, and Greytak}}]{fkw98}
\bibinfo{author}{\bibfnamefont{D.~G.} \bibnamefont{Fried}} \bibnamefont{{\it et al.}},
  \bibinfo{journal}{{ Phys. Rev. Lett.}} \textbf{\bibinfo{volume}{81}},
  \bibinfo{pages}{3811} (\bibinfo{year}{1998}).

\end{thebibliography}

\end{document}